\documentclass[10pt,amsmath,amssymb]{article}
\usepackage{amssymb,amsfonts}
\font\mybb=msbm10 at 12pt
\def\bb#1{\hbox{\mybb#1}}
\begin{document}
\begin{flushright}
arXiv:0912.1741 [hep-th]  
\end{flushright} \vspace{1.5cm}

\begin{center}
{\LARGE\bf On multiple M2-brane model(s) and its ${\cal N}$=8 superspace
formulation(s) }\footnote{Contribution to the {\it  Selected Topics in Mathematical and Particle
Physics Prague 2009}. Procs. of the  Conference on the occasion of Prof. Jiri
Niederle's 70th birthday. }

\vspace{1.1cm}

{\large\bf Igor A. Bandos}

\vspace{0.8cm}

{\it Ikerbasque, the Basque Foundation for Science and \\ Department
of Theoretical Physics,
 University of the Basque Country, \\ P.O. Box 644,
48080 Bilbao, Spain }\footnote{Also at the A.I. Akhiezer Institute
for Theoretical Physics, NSC KIPT,
Kharkov, Ukraine. E-mail: igor\_bandos@ehu.es}

\end{center}

{\small We give a brief review of Bagger-Lambert-Gustavsson (BLG)
model, with emphasis on its version invariant under the volume
preserving diffeomorphisms (SDiff$_3$) symmetry. We describe the
on-shell superfield formulation of this SDiff$_3$ BLG model in standard ${\cal N}=8$, d=3
superspace,  as well as its superfield action in the pure spinor ${\cal N}=8$ superspace. We also briefly address the Aharony-Bergman-Jafferis-Maldacena (ABJM/ABJ) model invariant under
$SU(M)_{k}\times SU(N)_{-k}$ gauge symmetry,  and discuss the possible form of their ${\cal N}=6$ and, for the case of Chern-Simons level $k=1,2$, ${\cal N}=8$ superfield equations.}

\vspace{0.8cm}

\section{Introduction. }
\setcounter{equation}{0}

In the fall of 2007, motivated by the search for a low-energy
description of the multiple M2-brane system, Bagger, Lambert and
Gustavsson \cite{Bagger:2007jr,Bagger:2007vi,Gustavsson:2007vu}
proposed a ${\cal N}=8$ supersymmetric superconformal $d=3$ model
based on Filippov three algebra \cite{Filippov} instead of Lie
algebra.
\medskip

{\bf 1.1. 3-algebras.} Lie algebras  are defined with the use of
antisymmetric brackets $[X, Y]=-[Y,X]$  of two elements,
$X=\sum\limits_a X^a T_a$ and $Y=\sum\limits_a Y^a T_a$, called Lie
brackets or commutator. The brackets of two Lie algebra generators,
$\; {}[T_a,T_b]=f_{ab}{}^cT_c$, are characterized  by antisymmetric
structure constants $f_{ab}{}^c=-f_{ab}{}^c= f_{[ab]}{}^c$  which
obey the Jacobi identity $ f_{[ab}{}^d f_{c]d}{}^e=0\;
\Leftrightarrow \;$ $\; [T_a,[T_b,T_c]]+ [T_c,[T_a,T_b]]+
[T_b,[T_c,T_a]]=0$.

In contrast, the general Filippov 3-algebra is defined by {\it
3-brackets}
\begin{eqnarray}\label{3balg-ra}
\{ T_a \, , \, T_b \, , \, T_c \} =  f_{abc}{}^{d}\, T_d
\; ,  \qquad f_{abc}{}^d = f_{[abc]}{}^d \; \qquad
\end{eqnarray}
which are antisymmetric and obey the so-called `fundamental
identity'
\begin{eqnarray}\label{Fid3alg} \{ T_a, T_b, \{ T_{c_1}, T_{c_2}, T_{c_3}\}\} = 3 \{ \{ T_a, T_b, T_{[c_1}\} , T_{c_2}, T_{c_3]}\}\}\; .  \qquad
\end{eqnarray}
To write an action for some 3-algebra valued field theory, one needs as well to introduce an invariant
inner product or metric
\begin{eqnarray}\label{hab=3-alg} h_{ab}=<T_a,T_b>\; .  \qquad
\end{eqnarray}  Then for the
metric 3-algebra the structure constants obey
$f_{abcd}:=f_{abc}{}^eh_{ed}=f_{[abcd]}$.

An example of infinite dimensional 3-algebra is defined by the Nambu
brackets (NB) \cite{Nambu:1973qe} of functions on a 3-dimensional
manifold $M^3$
\begin{eqnarray}\label{Nambu3br}
\{ \Phi , \Xi , \Omega \} =  \epsilon^{ijk}\; \partial_i \Phi \;
\partial_j \Xi \; \partial_k \Omega\; , \qquad \partial_i:=
\partial/\partial y^i\; , \qquad i=1,2,3\; . \qquad
\end{eqnarray}
Here $y^i=(y^1,y^2, y^3)$ are local coordinates on $M^3$, $\Phi =
\Phi (y)$,  $\Xi = \Xi (y)$ and $ \Omega= \Omega (y)$ are functions
on $M^3$,  and $\epsilon^{ijk}$ is the Levi-Cevita symbol (it is convenient to define NB  using a constant scalar density $e$ \cite{Bandos:2008fr}, but this is not important for our present discussion here and we simplify the notation by setting $e=1$). These brackets are invariant with respect to
the {\it volume preserving diffeomorphisms of $M_3$}, which we call
{\it SDiff$_3$ transformations}. In practical applications one needs to assume
compactness of $M^3$. For our discussion here it is
sufficient to assume that $M^3$ has the topology of sphere $S^3$.

Another example of 3-algebra, which was present already in the first
paper by Bagger and Lambert \cite{Bagger:2007jr} is ${\cal A}_4$
realized by generators $T_a$, $a=1,2,3,4$ obeying
\begin{eqnarray}\label{3balA4}
\{ T_a \, , \, T_b \, , \, T_c \} =  \epsilon_{abcd}\, T_d \; ,
\qquad a, b,c,d =1,2,3,4\; . \qquad
\end{eqnarray}
These are related to the 6 generators $M_{ab}$ of $SO(4)$ as
Euclidean $d=4$ Dirac matrices are related to the
$Spin(4)=SU(2)\times SU(2)$ generators, $T_a\leftrightarrow
\gamma_a$, $M_{ab} \leftrightarrow 1/2 \gamma_{ab}:= 1/4
(\gamma_a\gamma_b - \gamma_b\gamma_a)$.

A more general type of 3-algebras with not completely antisymmetric
structure constants were discussed {\it e.g. } in
\cite{Cherkis:2008qr}, \cite{Bagger:2008se} and
\cite{Gustavsson:2009pm}. In particular, as it was shown in
\cite{Bagger:2008se}, the Aharony-Bergman-Jafferis-Maldacena (ABJM)
model \cite{Aharony:2008ug} is based on a particular 'hermitian
3-algebra' the 3-brackets of which can be defined on two $M\times N$ (complex) matrices ${\bb Z}^i$, ${\bb Z}^j$ and an $N\times M$ (complex) matrix ${\bb Z}^\dagger_k$ by \cite{Bagger:2008se}
\begin{eqnarray}\label{h3br}
[ {\bb Z}^i\, , {\bb Z}^j\, ; {\bb Z}^\dagger_k ]^{^{M\times N}} =
{\bb Z}^i {\bb Z}^\dagger_k{\bb Z}^j - {\bb Z}^j {\bb Z}^\dagger_k{\bb Z}^i\; .   \qquad
\end{eqnarray}

\medskip

{\bf 1.2. BLG action.} The BLG model on general 3-algebra is described in terms of an octet of 3-algebra valued scalar fields
in vector representation of SO(8), $\phi^I(x)= \phi^{Ia}(x) T_a$, an octet of
3-algebra valued spinor fields in spinor (say, s-spinor) representation of SO(8), $\psi_{\alpha A}(x)= \psi_{\alpha A}{}^{a}(x) T_a$, and the vector gauge field $A_\mu^{ab}$ in the bi-fundamental representation of the 3-algebra. The BLG Lagrangian reads
\begin{eqnarray}\label{LagA4}
{\cal L}_{BLG} =Tr\, \left[ -\frac{1}{2} \left| {\cal D}\phi\right|^2 - \frac{g^2}{12}  \left\{\phi^I,\phi^J,\phi^K\right\}^2 - \frac{i}{2}  \bar\psi \gamma^\mu {\cal D}_\mu \psi  \right. \hspace{4cm} \qquad \\ \nonumber {}\hspace{4cm} \left. +  \frac{ig}{4} \left\{\phi^I\, , \phi^J \, , \bar\psi\right\} \rho^{IJ}\psi
\right]
+ \ \frac{1}{2g}{\cal L}_{CS}\, , \qquad I=1,\ldots , 8\; . \quad
\end{eqnarray}
where $g$ is a real dimensionless parameter,  ${\cal L}_{CS}$ is the Chern-Simons (CS term) for the gauge potential $A_{\mu}{}_b{}^a= A_{\mu}^{cd}f_{dcb}{}^a$ which is also
used to define the covariant derivatives of the scalar and spinor
fields. The $Spin(8)$ indices are suppressed in (\ref{LagA4});  $\rho^I:= \rho^I_{A\dot{B}}$ are the $8\times 8$ $Spin(8)$ `sigma' matrices (Klebsh-Gordan coefficients relating the  vector
{\bf 8}$_{v}$ and two spinor, {\bf 8}$_{s}$ and {\bf 8}$_{c}$,
representations of $SO(8)$). These  obey $\rho^I\tilde{\rho}{}^J+
\rho^I\tilde{\rho}{}^J= 2\delta^{IJ} \, I$ with their transpose
$\tilde\rho^I:= \tilde\rho^I_{\dot{A}B}$; notice that $\rho^{IJ} :=
(\rho^{[I}\tilde\rho^{J]})_{AB}$ and $\tilde{\rho}{}^{IJ} :=
(\tilde\rho^{[I}\rho^{J]})_{\dot A\dot B}$ are {\it antisymmetric}
in their spinor indices.

This model possesses ${\cal N}=8$ supersymmetry and superconformal
symmetries the set of which includes 8 special conformal
supersymmetries.  Hence the total number of supersymmetry parameters
is 2$\times$8+2$\times$8=32. This coincides with the number of supersymmetries possessed by  M2-brane \cite{Bergshoeff:1987cm} and the conformal symmetry was expected for {\it
infrared fixed point} (low energy approximation) of the multiple
M2-brane system \cite{Bandres:2008vf}. Thus, action
(\ref{LagA4}) was expected to play for the multiple M$2$-brane system
the same r\^ole as it is played by the $U(N)$ SYM action for the
multiple D$p$-brane system  \cite{Witten:1995im} (with N D$p$-branes).

However, if this were the case, the number of generators of the
Filippov 3-algebra would be related somehow to the number of M2-branes
composing the system the low energy limit of which  is described by
the action (\ref{LagA4}). This expectation enters in conflict with
the relatively poor structure of the set of finite dimensional
Filippov 3-algebras with positively definite metric (\ref{hab=3-alg}): this
set was proved to contain the  direct sums of ${\cal A}_4$ and
trivial one-dimensional 3-algebras only (see
\cite{Papadopoulos:2008sk,Gauntlett:2008uf} as well as
\cite{deAzcarraga:2009pr} and refs therein).

A very useful r\^ole in searching for resolution of this paradox was played by the
analysis by Raamsdock \cite{VanRaamsdonk:2008ft}, who reformulated
the ${\cal A}_4$ BLG model in matrix notation. This was used by
Aharony, Bergman, Jafferis and Maldacena \cite{Aharony:2008ug} to
formulate an $SU(N)_k\times SU(N)_{-k}$ and then
\cite{Aharony:2008gk} $SU(M)_{k}\times SU(N)_{-k}$ gauge invariant
CS plus matter models, which are believed to describe the low energy multiple
M2-brane dynamics. The subscript $k$ denotes the so-called {\it
CS level}, this is to say the integer coefficient in front of the CS
term in the action of the CS plus matter models. In the dual description of the ABJM model by  M-theory on the $AdS_4\times S^7/{\bb Z}_k$ \cite{Aharony:2008ug} the same integer $k$ characterizes the quotient of the 7-sphere.

The ABJM/ABJ model possesses only ${\cal N}=6$ manifest supersymmetries, which is natural for $k>2$, as the $AdS_4\times S^7/{\bb Z}_k$ backgrounds with $k>2$ preserve only $24$ of $32$ M-theory supersymmetries in these cases. The nonperturbative restoration of ${\cal N}=8$ supersymmetry for $k=1,2$ cases was conjectured already in \cite{Aharony:2008ug}. Recently this enhancement of supersymmetry was studied in \cite{Gustavsson:2009pm}, where its relation with some special `identities' (which we propose to call GR-identities or Gustavsson--Rey identities) conjectured to be true due to the properties of  monopole operators specific for $k=1,2$ is proposed. We shortly discuss the ABJM/ABJ  model in the concluding part of this paper.

\medskip {\bf 1.3. NB BLG action.} Coming back to the 3-algebra BLG models, we notice that inside their set there are clear candidates for the $N\to\infty$ limit of the multiple M2-brane system, which one can view as describing possible `condensates' of coincident planar M2-branes. These are the BLG theories in which the Filippov 3-algebra is realized by the Nambu-bracket (\ref{Nambu3br}) of functions defined on some 3-manifold $M_3$. This model was conjectured \cite{Ho:2008nn,Ho:2008ve} to be related with the
M5-brane \cite{Howe:1996yn,Bandos:1997ui,Schwarz+m5} wrapped over $M_3$ (see \cite{Bandos:2008fr} and recent \cite{Pasti:2009xc} for further
study of this proposal) and was put in a general context of
SDiff$_3$ gauge theories in \cite{IB+PKT=08-2}.

It is  described in terms of $Spin(8)$ ${\bf 8}_v$-plet of real
scalar fields $\phi^I$ ($I=1,\dots 8$), and a $Spin(8)$ ${\bf
8}_s$-plet of Majorana anticommuting $Sl(2;{\bb R})$ spinor fields
$\psi_A$ ($A=1,\dots 8$), both on the Cartesian product of
3-dimensional Minkowski spacetime with some $3$-dimensional closed
manifold without boundary, $M_3$. These fields transforms as scalars with respect to SDiff$_3$: $ \delta_\xi \phi = - \xi^i\partial_i \phi\,$, $ \delta_\xi \psi = - \xi^i\partial_i
\psi\,$, where $\xi^i=\xi^i(y)$ is a divergenceless SDiff$_3$ parameter.

The action of this Nambu bracket
realization of the Bagger--Lambert--Gustavsson model (NB BLG model)
is \begin{eqnarray}\label{Lagzero} {\cal L}_{NB\; BLG} &=& \oint
\!d^3y\, \left[ -\frac{1}{2}e \left| {\cal D}\phi\right|^2 -
\frac{i}{2} e\, \bar\psi \gamma^\mu {\cal D}_\mu \psi + \frac{ig}{4}
\varepsilon^{ijk}
\partial_i\phi^I\partial_j\phi^J \left(\partial_k\bar\psi
\rho^{IJ}\psi\right) \right. \nonumber \\  &&\qquad  \qquad \left. -
\frac{g^2}{12} e \left\{\phi^I,\phi^J,\phi^K\right\}^2 \right] + \
\frac{1}{2g}{\cal L}_{CS}\,  \qquad
\end{eqnarray}
In (\ref{Lagzero}) the trace $Tr$ of (\ref{LagA4})
is replaced by integral $\oint \!d^3y\,$ over  $M_3$ and ${\cal L}_{CS}$ is the CS-like
term involving the SDiff$_3$ gauge potential $s_i$ and gauge
pre-potential $A_i$ \cite{IB+PKT=08-2}. The gauge potential $s^i=dx^\mu s_\mu^i$ transforms under the local SDiff$_3$ with $\xi^i=\xi^i(x,y)$ as $\delta_\xi s^i = d\xi^i - \xi^j \partial_j s^i +  s^j \partial_j
\xi^i$ and is used to construct SDiff$_3$ covariant derivatives of scalar and spinor fields
\begin{eqnarray} {\cal D}\phi=d\phi + s^i\partial_i\phi\, , \qquad
{\cal D}\psi=d\psi + s^i\partial_i\psi\,  .
\end{eqnarray}
As the gauge field takes values in the Lie algebra of the Lie group
of gauge symmetries, and this is associated with volume
preserving diffeomorphisms the infinitesimal parameter of which is a
divergenceless three-vector $\xi^i(x,y)$, $\partial_i\xi^i=0$, the
SDiff$_3$ gauge field $s^i=dx^\mu s_{\mu}^i(x,y)$ obeys
\begin{eqnarray}\label{con-s}
\partial_is^i  \equiv 0\, \qquad \Leftrightarrow \qquad \partial_is_\mu^i  \equiv 0\, \qquad
\end{eqnarray}
which implies the possibility to express it, at least locally, in
terms of gauge pre-potential one-form $A_i=dx^\mu A_{\mu i}(x)$,
\begin{eqnarray}\label{s=edA}
s^i=\epsilon^{ijk}\partial_jA_k \qquad \Leftrightarrow \qquad
s_\mu^i=\epsilon^{ijk}\partial_jA_{\mu k} \; .
\end{eqnarray}
Also the covariant field strength
\begin{eqnarray}\label{fieldstrength} F^i = ds^i + s^j\partial_j
s^i\, ={1\over 2}dx^\mu \wedge dx^\nu F_{\nu\mu}^i  \; . \qquad
\end{eqnarray} satisfies the additional identity
\begin{eqnarray}\label{divfree}
\partial_i  F^i\equiv 0\, \qquad \Leftrightarrow
\qquad \partial_i  F_{\mu\nu}^i\equiv 0\,
\end{eqnarray}
and can be expressed (locally)  in terms of pre-field strength,
\begin{eqnarray}\label{F=edG} F^i= \varepsilon^{ijk} \partial_j G_{k}\, \qquad \Leftrightarrow
\qquad F_{\mu\nu}^i= \varepsilon^{ijk} \partial_j G_{{\mu\nu}\, k}\, , \qquad \\
\label{G3} G_i =dA_i
+ s^j\partial_{[j}A_{i]}={1\over 2}dx^\mu \wedge dx^\nu G_{\nu\mu i}
\, .
\end{eqnarray} The CS--like term in (\ref{Lagzero}) is expressed
through the gauge potential and pre-potential by
 \begin{eqnarray}\label{LCS=SDiff} {\cal L}_{CS} =
\oint d^3y \,  \epsilon^{\mu\nu\rho} \left[ \left(\partial_\mu
s_\nu^i\right) A_{\rho\, i} - \frac{1}{3} \epsilon_{ijk} s_\mu^i
s_\nu ^j s_\rho^k\right] \, , \qquad  \end{eqnarray} or, in
terms of differential forms, by $L_{CS}  =  \oint \! d^3y\,
\left[ds^i \wedge A_i - \frac{1}{3} \epsilon_{ijk}s^i \wedge
s^j\wedge s^k\right]$. The formal exterior derivative of $L_{CS}$
can be expressed through the field strength and pre-field strength
by
\begin{eqnarray}  d L_{CS} = \oint d^3y \,
F^i \wedge G_i \; .  \qquad
 \end{eqnarray}

The Lagrangian density (\ref{Lagzero}) varies into a total spacetime
derivative under the following infinitesimal supersymmetry
transformations with ${\bf 8}_c$-plet constant anticommuting spinor
parameter $\epsilon^\alpha_{\dot A}$ ($\dot A=1,\dots,8$):
\begin{eqnarray}\label{trans}
& \delta\phi^I = i \epsilon \tilde\rho^I \psi\, , \quad
\delta A_{\mu i} = -ig \left(\epsilon\gamma_\mu \tilde\rho^I\psi\right) \partial_i\phi^I\, , \quad
\delta\psi  = \left[\gamma^\mu  \rho^I {\cal D}_\mu \phi^I -
 \frac{g}{6} \left\{\phi^I,\phi^J,\phi^K\right\} \rho^{IJK}\right] \epsilon\, .\quad
\end{eqnarray}

The BLG equations of motion are
\begin{eqnarray}\label{EofM}
 {\cal D}^\mu {\cal D}_\mu \phi^I &=& {ig\over 2} \,
\varepsilon^{ijk}\partial_i\phi^J \partial_j\bar\psi \rho^{IJ}
\partial_k \psi -
{g^2\over 2}\left\{\left\{\phi^I,\phi^J,\phi^K\right\},
\phi^J,\phi^K\right\} \, ,  \nonumber \\
 \gamma^\mu
{\cal D}_\mu \psi &=& - {{g}\over 2} \; \rho^{IJ}
\left\{\phi^I,\phi^J,\psi\right\}\, ,  \\
 F_{\mu\nu}^i  &=&-  g \,
\varepsilon_{\mu\nu\rho} \varepsilon^{ijk}\left[ \partial_j\phi^I {\cal D}^\rho
\partial_k \phi^I -\frac{i}{2} \partial_j\psi\gamma^\rho
\partial_k\psi\right] \, . \nonumber
\end{eqnarray}

\section{NB BLG in ${\cal N}=8$ superfields}
\label{subsec:on-shell}

The NB BLG equations of motion can be obtained from the set of superfield equations in ${\cal N}=8$ superspace \cite{Bandos:2008df}. We will review this approach in this section.

Let us introduce ${\bf 8}_v$-plet of scalar, and SDiff$_3$-scalar,
superfields $\phi^I$, the lowest component of which (also denoted by
$\phi^I$) may be identified with the BLG scalar fields, and  impose
on it the following {\it superembedding--like} equation
\cite{Bandos:2008df}\footnote{The name comes from the observation that (\ref{DX=rp}) can be obtained from the {\it superembedding equation} for a single M2--brane \cite{bpstv} by first linearizing with respect to the dynamical fields in the static gauge, and then covariantizing the result with respect to SDiff$_3$.} \begin{eqnarray}\label{DX=rp} \fbox{$\; \bb{D}_{\alpha
\dot{A}} \phi^I= i \tilde{\rho}^I_{\dot{A}B} \psi_{\alpha B}\;$ }\,
. \qquad \end{eqnarray}

The SDiff$_3$-covariant spinorial derivatives on ${\cal N}=8$ superspace,
entering (\ref{DX=rp}),
\begin{eqnarray}\label{bbDs}
 \bb D_{\alpha\dot A} = D_{\alpha\dot
A} + \varsigma_{\alpha\dot A}{}^i
\partial_i \; , \qquad
\qquad \end{eqnarray} are constrained to obey the following algebra
\cite{Bandos:2008df}
\begin{eqnarray}\label{D21}
{} [\bb{D}_{\alpha \dot{A}} , \bb{D}_{\beta \dot{B}}]_+  &=& 2i
\delta_{\dot{A}\dot{B}} (C\gamma^\mu)_{\alpha\beta}{\cal D}_\mu + 2i
\epsilon_{\alpha\beta} W_{\dot{A}\dot{B}}{}^i\,
\partial_i \; , \quad
\end{eqnarray}
where  ${\cal
D}_\mu=\partial_\mu + is_\mu^i\partial_i$ is the 3-vector covariant
derivative which obeys
\begin{eqnarray}
 \label{D21+}
\left[{\bb D}_{\alpha \dot A}, {\cal D}_\mu \right] &=& F_{\alpha \dot A\, \mu}{}^i
\partial_i  \; , \qquad
\left[{\cal D}_\mu, {\cal D}_\nu \right] = F_{\mu\nu}{}^i \,
\partial_i\, . \qquad
\end{eqnarray}
Eqs. (\ref{D21}), (\ref{D21+}) are equivalent to the Ricci identity
${\cal D}{\cal D}={
 F}^i\partial_i$ for the covariant exterior derivative $ {\cal D} := d+ s^i\partial_i
= E^{\alpha \dot A} {\bb D}_{\alpha \dot A} + E^\mu {\cal D}_\mu\,
$, plus the constraint $ F^i_{\alpha \dot{A}\; \beta \dot{B}} =
2iC_{\alpha\beta} \, W_{\dot{A}\dot{B}}{}^i$.

The basic SDiff$_3$ gauge superfield strength
$W_{\dot{A}\dot{B}}{}^i$ is antisymmetric on c-spinor indices (this
is to say $W_{\dot{A}\dot{B}}{}^i$ is in the {\bf 28} of SO(8)); it
is also divergence-free, so
\begin{eqnarray}
W_{\dot{A}\dot{B}}{}^i=-W_{\dot{B}\dot{A}}{}^i\, , \qquad
\partial_i W_{\dot{A}\dot{B}}{}^i=0\, .
\end{eqnarray}
 Using the Bianchi identity $DF^i=0$,
 one finds that
\begin{eqnarray}
 F_{\alpha \dot A\, \mu}{}^i  = i  \left(\gamma_\mu W_{ \dot A}{}^i \right)_\alpha\, , \qquad
  W_{\alpha \dot B} {}^i := \frac{i}{7} {\bb D}_{\alpha \dot A} W_{\dot A\dot B}{}^i \,
  , \qquad
   F_{\mu\nu}{}^i = \frac{1}{16}
\epsilon_{\mu\nu\rho} {\bb D}_{\dot A} \gamma^\rho W_{\dot A}{}^i \,
, \qquad \end{eqnarray} and that
\begin{eqnarray}
 \label{DW=WI}
 {\bb{D}}_{\alpha (\dot A} W_{\dot{B} )\dot C}{}^i  =  i W_{\alpha \dot D}{}^i
 \left(\delta_{\dot D (\dot A} \delta_{\dot B )\dot C} - \delta_{\dot D\dot C}\delta_{\dot A\dot B}\right)\,
 ,
\qquad \\
\label{DW=DW+F}
 \bb{D}_{\dot{A} \alpha} W_{\beta\dot{B}}{}^i  =
(C\gamma^\mu)_{\alpha\beta}\left( {\cal D}_\mu
W_{\dot{A}\dot{B}}{}^i - 4 \delta_{\dot{A}\dot{B}} W_\mu
{}^i\right)\, .  \qquad  \end{eqnarray}
We see that the SDiff field strength supermultiplet includes a scalar  ${\bf 28}$ ($W_{\dot
A\dot B}{}^i$), a spinor ${\bf 8}_c$ ($W_{\alpha\dot A}{}^i$) and a
singlet divergence-free vector ($W^\mu{}^i= {\bb D}_{\dot A}
\gamma^\rho W_{\dot A}{}^i$).  There are many other independent
components, but these become dependent on-shell as far as we are
searching for a description of Chern--Simons (CS) rather than the
Yang--Mills one. The relevant super-Chern--Simons (super-CS) system
superfield equation {\it in the absence of `matter' supermutiplets}
is obviously $W_{\dot A\dot B}{}^i=0$, since this sets to zero all
SDiff$_3$ field strengths; in particular it implies $F_{\mu\nu}^i=0$.
In the presence of matter, the super-CS equation may get a nonvanishing right hand side.

Indeed, acting on the superembedding--like equation (\ref{DX=rp})
with an SDiff$_3$-covariant spinor derivative, and making use of the
anticommutation relation (\ref{D21}), one finds that $
\bb{D}_{\alpha [\dot{A} }\tilde{\rho}^{I}{}_{\dot{B}]C}
\psi^\alpha_C = 2 W_{\dot{A}\dot{B}}{}^i\partial_i\phi^I\, $ which
is solved by   the `super-CS' equation \cite{Bandos:2008df}
\begin{eqnarray}\label{WdAdBi=} \fbox{$\; W_{\dot{A}\dot{B}}{}^i=
 {2 g }\varepsilon^{ijk}\partial_i\phi^I\partial_j\phi^J\tilde{\rho}^{IJ}_{\dot{A}\dot{B}}\;$} \, .
\end{eqnarray}
It was shown in  \cite{Bandos:2008df} that  the two ${\cal N}=8$
superfield equations (\ref{DX=rp}) and (\ref{WdAdBi=}) imply the
Nambu-bracket BLG equations (\ref{EofM}).

\section{NB BLG in pure-spinor superspace}

An  ${\cal N}=8$ superfield action for the abstract BLG model, {\it i.e.} for the BLG model based on a finite dimensional 3-algebra, which in practical terms implies ${\cal A}_4$ or the direct sum of several ${\cal A}_4$ and trivial 3-algebras, was proposed by  Cederwall \cite{Cederwall:2008vd}. Its generalization for the case of NB BLG model invariant under infinite dimensional SDiff$_3$ gauge symmetry, constructed in \cite{IB+PKT=08-2}, will be reviewed in this section.

The pure-spinor superspace of \cite{Cederwall:2008vd} is parametrized by the standard ${\cal N}=8$
$D=3$ superspace coordinates $(x^\mu ,\theta^\alpha_{\dot{A}})$ together with additional pure spinor coordinates $\lambda^\alpha_{\dot A}$. These are described by the ${\bf 8}_c$-plet of complex commuting $D=3$ spinors satisfying the `pure spinor' constraint
\begin{eqnarray}\label{N8pure} \lambda \gamma^{\mu}
\lambda:=\lambda^{\alpha}_{\dot A}
\gamma^{\mu}_{\alpha\beta} \lambda^{\beta}_{\dot A} =0\, . \qquad
\end{eqnarray}

This is a variant of the $D=10$ pure-spinor superspace first proposed by Howe \cite{Howe:1991mf} (see \cite{Nilsson:1985cm} for earlier attempt to use pure spinors in the SYM and supergravity context).
{} From a more general perspective, the approach of \cite{Cederwall:2008vd} can be considered as a realization of the harmonic superspace programme of \cite{Galperin:1984av} (although one cannot state that the algebra of all the symmetries of the superfield action of \cite{Cederwall:2008vd} are closed off shell, {\it i.e.} without the use of equations of motion).
The $D=10$ pure spinors are also the central element of the Berkovits approach to covariant description of quantum superstring \cite{Berkovits}. In this approach the pure spinors are considered to be the ghosts of a local fermionic gauge symmetry related to the $\kappa$--symmetry of the standard Green--Schwarz formulation. This `ghost nature' may be considered as a justification for that the pure-spinor superfields are assumed (in \cite{Cederwall:2008vd,IB+PKT=08-2} and here) to be {\it analytic} functions of $\lambda$ that can be expanded as a Taylor series in powers of $\lambda$. To discuss the BLG model, we allow all the pure spinor superfields to depend also on the local coordinates $y^i$ of the auxiliary compact 3-dimensional manifold $M_3$.

Following \cite{Cederwall:2008vd}, we define the BRST-type operator ({\it cf.} \cite{Berkovits})
\begin{eqnarray}\label{Q=BRST} Q := \lambda^\alpha_{\dot A} D_{\alpha\dot A}\, ,
\end{eqnarray} which satisfies $Q^2\equiv 0$ as a consequence of the
pure spinor constraint (\ref{N8pure}). We now introduce  the ${\bf 8}_v$-plet of  complex scalar ${\cal N}=8$ `matter' superfields $\Phi^I$, with SDiff$_3$ transformation
 \begin{eqnarray}\label{SDiff=Phi}
 \delta\Phi^I = \Xi^i\partial_i \Phi^I\,
 \end{eqnarray}
characterized by the commuting $M_3$-vector parameter $\Xi^i= \Xi^i (y)$.

We allow these superfields  to be complex because they may depend on
the complex  pure-spinor $\lambda$  but, to make contact with the spacetime BLG model, we
assume that the leading term in its decomposition in power series on complex $\lambda$
\begin{eqnarray} \Phi^I = \phi^I + {\cal O}\left(\lambda\right) \, ,
 \end{eqnarray}
is given by a {\it real} ${\bf 8}_v$-plet of  `standard'  ${\cal N}=8$
scalar superfields, like the basic objects in Sec. \ref{subsec:on-shell}.

Let us consider (complex and anticommuting) Lagrangian density
\begin{eqnarray}\label{L-matt0}
{\bb L}^0_{mat} = {1\over 2}M_{IJ} \oint d^3y \, e \Phi^I  Q
\Phi^J\, ,
\end{eqnarray}
where $M_{IJ}= {\lambda}_{\dot{A}}^\alpha\,
\tilde{\rho}^{IJ}_{\dot{A}\dot{B}}\lambda_{\alpha\dot{B}}$ is one of
the two  nonvanishing analytic pure spinor bilinears
\begin{eqnarray}\label{bilinears}
 M_{IJ}:= {\lambda}^\alpha\, \tilde{\rho}^{IJ}\lambda_\alpha\, , \qquad
 N^\mu_{IJKL}:= {\lambda}\, \gamma^\mu \tilde{\rho}^{IJKL}\lambda\, .
 \end{eqnarray}
It is important that, due to (\ref{N8pure}), these obey the identities (see \cite{IB+PKT=08-2} for a detailed
proof) \begin{eqnarray}\label{IDone} M_{IJ} \, \tilde\rho^J\lambda \equiv
0\, , \qquad \ M_{[IJ} M_{KL]}=0 \, , \qquad  \ N_{PQ[IJ} \cdot
N_{KL]PQ} \equiv 0\, .  \end{eqnarray}

To construct the ${\cal N}=8$ supersymmetric action with the use of the Lagrangian (\ref{L-matt0}) one needs to specify an adequate superspace integration measure. We refer to \cite{Cederwall:2008xu}
for details on such a measure, which has the crucial property of allowing us to discard a BRST-exact terms when varying with respect $\Phi^I$. Then, as a consequence of this and also of the identities
(\ref{IDone}), the action is invariant under the gauge symmetries $
\delta \Phi^I= {\lambda}^\alpha_{\dot{A}}\tilde{\rho}_{\dot{A}B}^I \zeta_{\alpha B} + {Q} K^I $ for arbitrary pure-spinor-superfield parameters $\zeta_\alpha$ and
$K^I\;$.

The variation with respect to $\Phi^I$ yields the superfield equation
\begin{eqnarray}\label{mattereq=Q} M_{IJ} Q\Phi^J =0\, ,
 \end{eqnarray}
 which implies, as a consequence of the pure-spinor identities, that
 \begin{eqnarray}\label{QP=lrIT} Q\Phi^I =\lambda \tilde \rho^I \Theta
 \end{eqnarray}
for some ${\bf 8}_s$-plet of complex spinor superfields
$\Theta_{\alpha \dot A}$. The first nontrivial  ($\sim\lambda$) term
in the $\lambda$-expansion of this equation is  precisely the free
field limit of the on-shell superspace constraint (\ref{DX=rp}),
${D}_{\alpha \dot{A}} \phi^I= i \tilde{\rho}^I{}_{\dot{A}B}
\psi_{\alpha B}$,  with $\psi=\Theta|_{\lambda=0}$. \footnote{Notice
that the above mentioned  gauge symmetry $ \delta \Phi^I=
{\lambda}^\alpha_{\dot{A}}\tilde{\rho}_{\dot{A}B}^I \zeta_{\alpha
B}$ of the action (\ref{L-matt0}) contributes to $\delta
(Q\Phi^I)$ the terms of at least the second order in $\lambda$. Then
the induced transformation of the pure spinor superfield
$\Theta_{\alpha\dot{A}}$ in (\ref{QP=lrIT}) is of the first order in
$\lambda$ so that $\psi_{\alpha\dot{A}}=
\Theta_{\alpha\dot{A}}\vert_{\lambda =0 }$, entering the
superembedding-like equation (\ref{DX=rp}), is inert under those
transformations. } In the light of the results of Sec. 2, this
implies that   the free field ($g\mapsto 0$) limit of the NB BLG
field equations (\ref{EofM}) can be obtained from the pure spinor
superspace action (\ref{L-matt0}).

Now, as the free field limit is reproduced, to construct the pure
spinor superspace description of the NB BLG system we need to
describe its gauge field (Chern-Simons) sector and to use it to
gauge the  SDiff$_3$ invariance. To this end, we introduce an
$M_3$-vector-valued complex {\it anticommuting} scalar $\Psi^i$ with
the SDiff$_3$ gauge transformations
 \begin{eqnarray}\label{QSDiffPsi=} \delta \Psi^i
= Q \Xi^i  + \Psi^j\partial_j \, \Xi^i - \Xi^j\partial_j \Psi^i\, ,
\qquad
\partial_i\Xi^i=0\,
\end{eqnarray}
involving the commuting $M_3$-vector parameter $\Xi^i= \Xi^i(x,\theta, \lambda ; y^j)$ and its  derivatives. In the present context, $\Psi^i$ will play the role of the SDiff$_3$ gauge potential.  We require that $\partial_i\Psi^i=0$ so that, locally on $M_3$,
 \begin{eqnarray}\label{Psi=eedPi}
 \Psi^i =
\varepsilon^{ijk}\partial_j \, \Pi_k\, ,
 \end{eqnarray}
where $\Pi_i$ is the  complex {\it anticommuting}, and spacetime
scalar, pre-gauge potential of this formalism.

Using $\Psi^i$ we can define an SDiff$_3$-covariant extension of $Q\Phi^I$ by
 \begin{eqnarray}\label{bbQ=Q+}
 \bb Q \Phi^I := Q\Phi^I + \Psi^i\partial_i \Phi^I \,
 \end{eqnarray}
and construct the generalization of (\ref{L-matt0}) invariant under {\it local} SDiff$_3$ symmetry (\ref{SDiff=Phi}), (\ref{QSDiffPsi=}):
 \begin{eqnarray}\label{L-matt} {\bb L}_{mat} = {1\over 2}M_{IJ} \oint d^3y \, e \Phi^I \bb Q
\Phi^J\, , \hspace{2cm} M_{IJ}\, = {\lambda}\,
\tilde{\rho}^{IJ}\epsilon \lambda\; .
\end{eqnarray}

Next we have to construct the (complex and fermionic) Lagrangian
density ${\bb L}_{CS}$ describing the (Chern-Simons) dynamics of the
gauge potential $\Psi^i$. To this end we introduce the field-strength superfield
 \begin{eqnarray}\label{FSS}
 {\cal F}^i := Q \Psi^i + \Psi^j
\partial_j \Psi^i\,  = \varepsilon^{ijk}\partial_j {\cal
G}_k\, ,
\end{eqnarray}
where the last equality is valid locally on $M_3$ and
\begin{eqnarray}{\cal G}_i  := Q\Pi_i + \Psi^j \partial_j \Psi_i\,
\end{eqnarray}
is the pre-field-strength superfield of this formalism. Both ${\cal
F}^i$ and ${\cal G}_i$ are SDiff$_3$ covariant, so ${\cal F}^i{\cal
G}_i$ is an SDiff$_3$ scalar. Furthermore, the  integral of this density over $M_3$ is
$Q$-exact, in the sense that
\begin{eqnarray} \int d^3y \, e\, {\cal F}^i  {\cal G}_i = Q \, \bb L_{CS}\, ,
\end{eqnarray}
where
\begin{eqnarray}{\bb L}_{CS} = \int d^3\sigma \, e\,  \left( \Pi_i \, Q\Psi^i -
\frac{1}{3}\epsilon_{ijk}\Psi^i\Psi^j\Psi^k \right)
\end{eqnarray}
is the complex and anti-commuting CS-type Lagrangian density
\cite{IB+PKT=08-2} which can be used, together with ${\bb L}_{mat}$
of (\ref{L-matt}),  to construct the candidate Lagrangian density of
the NB BLG model,
\begin{eqnarray}\label{matter+GF}
 {\bb L} = {\bb L}_{mat} - \frac{1}{g}{\bb L}_{CS}\; . \qquad
\end{eqnarray}

The $\Pi_i$ equation of motion of this combined  Lagrangian is
\begin{eqnarray}\label{calFeq} {\cal F}^i =  \frac{g}{2e}
M_{IJ}\epsilon^{ijk}\partial_j\Phi^I\partial_k\Phi^J \, .
\end{eqnarray} At this stage it is important to assume that $\Psi^i$
has `ghost number one'  \cite{Cederwall:2008vd}, which means that
it is a power series in $\lambda$ with vanishing zeroth order term
(and similarly for its pre-potential $\Pi_i$). In other words
\begin{eqnarray}\label{Psi=lsigma} \Psi^i = \lambda^\alpha_{\dot{A}}
\varsigma^i_{\alpha\dot{A}}\; ,
 \end{eqnarray}
where $\varsigma^i$ is an $M_3$-vector-valued  ${\bf 8}_c$-plet of
arbitrary anticommuting spinors. Its zeroth component  in the
$\lambda$-expansion is the fermionic  SDiff$_3$ potential
introduced, with the {\it same symbol}, in (\ref{bbDs}).  With this
`ghost number' assumption, (\ref{calFeq}) produces  at lowest
nontrivial order ($\sim \lambda^2$) the superspace constraints
(\ref{D21}) for the  `ghost number zero' contribution
$\varsigma^i\vert_{\lambda=0}$ to the pure spinor superfield
$\varsigma^i$ in (\ref{Psi=lsigma}), accompanied by the super CS
equation (\ref{WdAdBi=}) for the field strength $W_{\dot{A}\dot{B}}$
constructed from this potential.

An heuristic justification of the assumption (\ref{Psi=lsigma}), so
crucial to obtain the correct super-CS equations, can be found in
that with this form of $\Psi^i$ the covariantized BRST operator in
(\ref{bbQ=Q+}) does not contain a contribution of ghost number zero,
{\it i.e.} it has the form of (\ref{Q=BRST}), ${\bb Q}=
{\lambda}_{\dot{A}}{}^{\alpha}\; {\bb D}_{\alpha\dot{A}}$, but with
the SDiff$_3$ covariant Grassmann derivative ${\bb
D}_{\alpha\dot{A}}= D_{\alpha\dot{A}} +
\xi^i_{\alpha\dot{A}}\partial_i$.

Varying the interacting action with respect to $\Phi^I$ results in
SDiff$_3$ gauge invariant generalization of Eqs. (\ref{mattereq=Q}),
\begin{eqnarray}\label{mattereq} M_{IJ} \bb Q\Phi^J =0\, ,
 \end{eqnarray}
which contains, as the first nontrivial  ($\sim (\lambda )^3$)  term
in the $\lambda$-expansion,  precisely the superembedding--like
equation  (\ref{DX=rp}) with $\psi=\Theta|_{\lambda=0}$.

We have now shown, following \cite{IB+PKT=08-2}, how  the on-shell
${\cal N}=8$ superfield formulation of Sec. \ref{subsec:on-shell},
and hence all BLG field equations (\ref{EofM}), may be extracted
from the equations of motion derived from  the pure spinor
superspace action (\ref{matter+GF}). Of course, the field content
and equations of motion should  be analyzed at all higher-orders in
the $\lambda$-expansion. To this end, one must take into account the
existence of additional gauge invariance
\cite{Cederwall:2008vd,Cederwall:2008xu}
\begin{eqnarray}\label{QgaugePi=}
\delta \Phi^I= \bar{\lambda}\tilde{\rho}^I \zeta_\alpha + (\bb{Q}+
{\Psi}^j\; \partial_j) K^I\; ,  \qquad \delta {\Pi}_i = K^I\,
M_{IJ}\,
\partial_i\Phi^J\; ,
\end{eqnarray}
for arbitrary pure-spinor-superfield parameters $\zeta_\alpha$ and
$K^I\;$.

What one can certainly state, even without a detailed analysis of these symmetries, is that, if
additional fields are present inside the pure spinor superfields of the model (\ref{matter+GF}), they are decoupled from the BLG fields in the sense that they do not enter the equations of motion of the BLG fields which are obtained from the pure spinor superspace equations. This allowed us \cite{IB+PKT=08-2}, following the terminology of  \cite{Cederwall:2008vd}, to call (\ref{matter+GF}) the {\cal N}=8
superfield action for the NB BLG model.

\section{Remarks on ABJM/ABJ model}

 The ${\cal
N}=6$ pure spinor superspace action for the ABJM model
\cite{Aharony:2008ug} invariant under $SU(N)_k\times SU(N)_{-k}$
gauge symmetry, was proposed in
\cite{Cederwall:2008xu}\footnote{Notice the existence of the {\it
off-shell} ${\cal N}=3$ superfield formalism for the ABJM model
\cite{Buchbinder:2008vi} which was used to develop the quantum
calculation technique in \cite{Buchbinder:2009dc}}. One can extract
the standard (not pure spinor) ${\cal N}=6$ superspace equation by
varying the action of \cite{Cederwall:2008xu} and fixing its gauge
symmetries. It is also instructive (and probably simpler) to develop independently the
on-shell ${\cal N}=6$ superspace formalism for the ABJM as well as
for the ABJ \cite{Aharony:2008gk} model invariant under
$SU(M)_k\times SU(N)_{-k}$ symmetry \cite{IB+JdA=09}.

For any value of the CS-level $k$ the starting point of the on-shell
${\cal N}=6$ superfield formalism could be the following (superembedding-like) superspace
equation for complex $M\times N$ matrix superfield ${\bb Z}^i$
\cite{IB+JdA=09}\footnote{Here and below we use the Latin symbols
from the middle of the alphabet, $i,j,...$, to denote the
four-valued $SU(4)$ index, $i,j,...= 1,2,3,4$; we hope that this
will not produce  confusion with real 3-valued vector indices of
$M_3$, see secs. 1.3, 2 and 3, as far as we do not use these in the
present discussion.}
\begin{eqnarray}\label{DZi=6N}
 {\bb D}_\alpha^I {\bb Z}^i=\tilde{\gamma}^{I ij} {\psi}{}_{\alpha j}\; ,
 \qquad I=1,2,...,6 \; , \qquad i,j=1,2,3,4 \; . \qquad
\end{eqnarray}
Here $\tilde{\gamma}^{I ij} = {1\over 2}\epsilon^{ijkl}
{\gamma}^{I}_{kl}= -({\gamma}^{I}_{ij})^*$ and ${\gamma}^{I}_{ ij}= -
{\gamma}^{I}_{ ji}$ are $SO(6)$ Klebsh-Gordan coefficients (generalized Pauli
matrices), which obey
${\gamma}^{I}\tilde{\gamma}^J+{\gamma}^{J}\tilde{\gamma}^I=\delta^{IJ}$.
The matrix superfield ${\bb Z}^i$  carries $({\bf M},\bar{\bf N})$
representation of the $SU(M)\times SU(N)$ gauge group. Its hermitian
conjugate ${\bb Z}^\dagger_i$ is $N\times M$ matrix carrying
$(\bar{\bf M}, {\bf N})$ representation and obeying $ {\bb
D}_\alpha^I {\bb Z}{}^\dagger_i ={\gamma}^{I}_{ ij}
{\psi}{}_{\alpha}^{\dagger j}$. Notice that, although in the original
ABJM model \cite{Aharony:2008ug} $M=N$, the $N\times N$ matrix
superfields ${\bb Z}^i$ and ${\bb Z}^\dagger_i$ carry different
representation of $SU(N)\times SU(N)$: $({\bf N},\bar{\bf N})$ and
$(\bar{\bf N},{\bf N})$, respectively. Here we speak in terms of the
case with $M\not= N$, which is terminologically simpler, but all our
arguments clearly also apply for $M=N$.

The Grassmann spinorial covariant derivatives ${\bb D}_\alpha^I$ in
(\ref{DZi=6N}) includes the gauge group $SU(M)\times SU(N)$
connection and obey the algebra
\begin{eqnarray}\label{(DfI,DfJ)=6N}
{} && \{ {\bb D}_\alpha^I, {\bb D}_\beta^J\} = i
\gamma^a{}_{\alpha\beta} \delta_{IJ} {\cal D}_a + i
\epsilon_{\alpha\beta} W^{IJ} \; . \qquad \qquad
\end{eqnarray}
This algebra involves the $15$-plet of the basic field strength
superfields $W^{IJ}=-W^{JI}$ which can be expressed through the matter
superfields by  the following ${\cal N}=6$ super-CS  equation \cite{IB+JdA=09}
\begin{eqnarray}\label{sCS=6N}
 W_{SU(M)}^{IJ}=  i {\bb Z}^i {\bb Z}{}^\dagger_j\; {\gamma}^{IJ}{}_{i}{}^j
\; ,   \qquad W_{SU(N)}^{IJ}= i  {\bb Z}{}^\dagger_j {\bb Z}^i
{\gamma}^{IJ}{}_{i}{}^j  \; . \qquad
\end{eqnarray}
Here $W_{SU(M)}^{IJ}$ and $W_{SU(N)}^{IJ}$ are the basic field
strength corresponding to $SU(M)$ and $SU(N)$ subgroups of the gauge
group $SU(M)_k\times SU(N)_{-k}$. One can check that the consistency conditions for Eqs. (\ref{DZi=6N}) and  (\ref{sCS=6N}) are satisfied if the matter superfield obeys the superfield equation of motion
\begin{eqnarray}\label{sEqm=6Ns}
& \gamma^J_{ij} D^{\beta (I} D_{\beta}^{ J)} {\bb Z}^j + 4 \gamma^J_{ij} [ {\bb
Z}^j, {\bb Z}^k; {\bb Z}^\dagger_k]  + 3 \gamma^J_{jk} [ {\bb
Z}^j, {\bb Z}^k; {\bb Z}^\dagger_i]  =0\;  , \qquad
\end{eqnarray}
where $[ {\bb
Z}^j, {\bb Z}^k; {\bb Z}^\dagger_k]$ are hermitian 3--brackets (\ref{h3br}). This superfield equation  implies, in particular, the fermionic equations of motion \cite{IB+JdA=09}
\begin{eqnarray}\label{fEqm=6Ns}
& \gamma^a_{\alpha\beta} D_a\psi^\beta_i= i [\psi_{\alpha j}, {\bb Z}^j; {\bb Z}{}^\dagger_i] + {i\over 2} [\psi_{\alpha i}, {\bb Z}^j; {\bb Z}{}^\dagger_j] +{i\over 6}\epsilon_{ijkl} [{\bb Z}^j,  {\bb Z}^k; \psi_{\alpha}^{\dagger l}]   . \qquad
\end{eqnarray}
We refer to  \cite{IB+JdA=09} for further details on the ${\cal N}=6$ superspace formalism of the  ABJM/ABJ model, including for the explicit form of the bosonic equations of motion.

Searching for an ${\cal N}=8$ superfield formulation for the
ABJM/ABJ models with CS levels $k=1,2$ it is natural to assume that
the universal ${\cal N}=6$ sector is present as a part of ${\cal
N}=8$ superspace formalism and, to describe two additional fermionic directions of ${\cal
N}=8$ superspace, introduce, in addition
to six ${\bb D}_\alpha^I$, one complex spinor Grassmann  derivative
${\bb D}_\alpha$, and its conjugate $({\bb D}_\alpha )^\dagger = -
\bar{\bb D}_\alpha$ obeying
\begin{eqnarray}
\label{(Df,Df)=3d} {} && \{ {\bb D}_\alpha ,\bar{\bb D}_\beta\} = i
\gamma^a{}_{\alpha\beta} {\cal D}_a + i \epsilon_{\alpha\beta} W \;
, \qquad \{{\bb D}_\alpha, {\bb D}_\beta\} = 0 \; , \qquad  \{
\bar{\bb D}_\alpha , \bar{\bb D}_\beta\} = 0 \; , \qquad   \\
\label{(Df,bDf)=3d} {} && \{ {\bb D}_\alpha, {\bb D}_\beta^J\} = i
\epsilon_{\alpha\beta} W^{J} \; , \qquad  \{ \bar{\bb D}_\alpha ,
{\bb D}_\beta^J\} = i \epsilon_{\alpha\beta} \bar{W}^{J}  . \; \qquad
\end{eqnarray}
The structure of additional ${\cal N}=2$ supersymmetries proposed in \cite{Gustavsson:2009pm} suggests to impose on the basic {\cal N}=8 superfields the chirality condition in the new
fermionic directions \cite{IB+JdA=09},
\begin{eqnarray}\label{DZi=}
 \bar{\bb D}_\alpha {\bb Z}^i=0\; ,  \qquad  {\bb D}_\alpha {\bb Z}{}^\dagger_i= 0 \;   . \qquad
\end{eqnarray}
While the natural candidate for the super-CS equation for the $SO(6)$ scalar superfield strength
$W$ is \begin{eqnarray}\label{W8N=} W= {\bb Z}^i {\bb Z}{}^\dagger_i\;   , \qquad
\end{eqnarray} to write a possibly
consistent  super-CS equation for 6 complex field strength $W^{J}$, which has to be chiral,
${\bb D}_\alpha W^{J}=0=\bar{\bb D}_\alpha \bar{W}^{J}$, to provide the consistency of the constraints  (\ref{(Df,Df)=3d}), (\ref{(Df,bDf)=3d}) and (\ref{(DfI,DfJ)=6N}),
\begin{eqnarray}\label{sCS=8N}
&  \bar{W}_{SU(M)}^{J}= \propto  {\bb Z}^i {\gamma}^{J}_{ij}  \tilde{\bb Z}{}^j\; , \qquad
 {W}_{SU(M)}^{J}=  \propto \tilde{\bb Z}{}^\dagger_i\tilde{\gamma}^{J\, ij} {\bb Z}{}^\dagger_j  \; ,  \qquad
\end{eqnarray}
one needs to involve ''non-ABJM superfields'', the leading
components of which are the ''non-ABJM fields'' of
\cite{Gustavsson:2009pm}.  These are $N\times M$ matrix $\tilde{\bb
Z}^i$ and $M\times N$ matrix $\tilde{\bb Z}{}^\dagger_i$ which obey
\begin{eqnarray}\label{DtZi=}
 & \bar{\bb D}_\alpha \tilde{\bb Z}^i=0\; ,  \qquad  {\bb D}_\alpha \tilde{\bb Z}{}^\dagger_i= 0 \;   \qquad
\end{eqnarray}
and must be related with ABJM superfields ${\bb Z}^i$, ${\bb
Z}{}^\dagger_i$ by using the suitable monopole operators (converting
$(\bar{\bf M},{\bf N})$ representation into $({\bf M},\bar{\bf
N})$) which exist for the case of CS levels $k=1,2$ only
\cite{Gustavsson:2009pm}. According to \cite{Gustavsson:2009pm}, the existence of these monopole operators is reflected by the `identities' between hermitain three brackets (\ref{h3br}) of the ABJM and non-ABJM (super)fields. The set of these `GR--identities' includes
\begin{eqnarray}\label{GR=id}
 [ (...), \tilde{\bb Z}{}^{\dagger}_i \, ; \tilde{\bb Z}{}^i] \; &=& -
[ (...), {\bb Z}{}{}^i \; ;{\bb Z}{}^{\dagger}_i]\;  \;   . \qquad
\end{eqnarray}

The consistency of the system of ${\cal N}=8$ superfield equations
(\ref{DZi=6N})-- (\ref{sCS=8N}) and the set of GR--identities necessary for that are presently under investigation
\cite{IB+JdA=09}.

\bigskip

{\large\bf Acknowledgments.} The author thanks Jos\'e de Azc\'arraga,
Warren Siegel, Dmitri Sorokin, Paul Townsend and Linus Wulff for
useful discussions.  This work was partially supported by the
Spanish MICINN under the project FIS2008-1980 and by the
Ukrainian National Academy of Sciences and Russian RFFI grant
38/50--2008.

\bigskip

{\large\bf Notice added:} After this manuscript has been finished, a paper \cite{Samtleben:2009ts}
devoted to ${\cal N}=8$  superspace formulations of $d=3$ gauge theories appeared on the net.
It contains a detailed description of the on-shell ${\cal N}=8$ superspace formulation of the BLG model for finite dimensional three algebras, similar to the formulation of the SDiff$_3$ invariant Nambu bracket BLG model in \cite{Bandos:2008df}, and of its derivation starting from the gauge theory constraints and Bianchi identities. Also the component field content of the SYM model defined by the  constraints (\ref{D21}) and its finite-3-algebra counterpart  is discussed there.

\end{document}